\documentclass[12pt]{article}
\usepackage[dvips]{color}
\usepackage{epsfig}
\usepackage{amsmath}
\usepackage{graphicx}
\usepackage{color}

\begin{document}
\title{ Thermal Fluctuations in  a  Charged AdS Black Hole}
\author{{Behnam Pourhassan$^{a}$\thanks{Email:
b.pourhassan@du.ac.ir},\hspace{1mm} and Mir Faizal$^{b}$\thanks{Email:
f2mir@uwaterloo.ca}}\\
$^{a}${\small {\em  School of Physics, Damghan University, Damghan, 3671641167,  Iran}}\\
$^{b}${\small {\em Department  of  Physics and Astronomy, University of Waterloo,  Waterloo, ON N2L 3G1, Canada}}} \maketitle
\begin{abstract}
In this paper, we will analyze  the effects of thermal fluctuations on a  charged AdS black hole.
This will be done by analyzing the corrections to  black hole thermodynamics due  to these thermal fluctuations.
We will demonstrate that    the entropy of this black hole get corrected by 
logarithmic term. We will also calculate other corrections to other important thermodynamic quantities for this black hole.
Finally, we will use the corrected value of the specific heat to analyze the phase transition in this system.
\end{abstract}

\section{Introduction}
It has been known that if an entropy is not associated with a black holes, then the second law of thermodynamics will get violated 
\cite{1, 1a}. This was because  the absence of an entropy for a black hole can cause a spontaneous reduction in the entropy of the universe.
This will happen  when ever  an object with a finite entropy crossed the horizon.
Thus, the black holes are taken to be  maximum entropy objects, i.e.,
black holes have more entropy than any other object of  the same volume~\cite{2, 4}.
This maximum  entropy is proportional to the area of the black hole \cite{4a}, and this observation in turn has led to the
development of the holographic principle~\cite{5, 5a}. In fact,
almost all the approaches to quantum gravity predict the same relation between the area and the entropy of a black hole. This
relation is given by  $S = A/4$,
where $S$ is the entropy of the black hole and $A$ is the area.
However, this maximum entropy of the black holes
is expected to get corrected due to quantum fluctuations, and this is in turn expected to modify the holographic
principle \cite{6, 6a}. These quantum fluctuations become important as the black hole reduces its size by giving off
Hawking radiation. So, as the 
black holes get smaller in size due to the radiation of Hawking radiation, quantum fluctuations are expected to
correct the standard relation between the area and the entropy of a black hole. 

Such corrections have been evaluated using
several different approaches. The density of microstates for asymptotically flat black holes have been calculated
using the non-perturbative quantum   general
relativity  \cite{1z}. In this approach,  the   density of states has been obtained
by counting the conformal blocks of a well defined  conformal field theory.   It has thus been demonstrated that the
 non-perturbative quantum   general
relativity  does reproduce the Bekenstein entropy for the large black holes. However, it is also
possible to calculate the leading order corrections to this entropy using the  non-perturbative quantum   general
relativity, and it turns  a logarithmic correction. The Cardy formula has been used to argue that such logarithmic
corrections should occur in all the black holes whose microscopic degrees of freedom are described by a conformal field theory
\cite{card, ca}. Such logarithmic corrections have also been obtained for
three dimensional BTZ black holes by computing the exact
partition function \cite{gks}. Matter fields in
backgrounds of a black hole have been used to demonstrated the  leading order corrections to the
black hole entropy are logarithmic corrections \cite{other, other0, other1}.
It has also been demonstrated the string theoretical effects can generate logarithmic
corrections to the entropy of a black hole \cite{solo1, solo2, solo4, solo5}. 
In fact, the corrections to the  entropy of dilatonic black holes have been explicitly obtained \cite{jy}, 
and they again turn out to be logarithmic
corrections. Rademacher expansion of the partition function also has been used to
obtained such corrections to the entropy of a black hole \cite{bss}. 

It  may be know that recently it has been argued that the corrections to the black hole entropy can come from
Planckian deformation of space-time, due to the generalized uncertainty \cite{mi}.
It has been demonstrated
that such corrections can lead to the existence of black hole remnants, which can have important phenomenological
consequences \cite{r1}. It is expected that quantum fluctuations in the geometry will lead to thermal fluctuations in the
thermodynamics of the black holes.
It is possible to calculate the corrections to the entropy of a black hole  by considering the
thermal fluctuations around equilibrium \cite{l1, SPR}. The corresponding canonical ensemble is stable, as the
stability condition is equivalent to having a positive specific heat. 

It is possible to construct various asymptotically
AdS charged  black hole solutions  ~\cite{Lemos1,Lemos2, Lemos3, Huang,Aminneborg,brill,Mann,Anderson}.
In fact, it is possible to define
a  thermodynamic equilibrium, and thus obtain many
related  results  for such black holes \cite{01, 04, 05, 02}. So, it is interesting to analyze the effect of thermal fluctuations 
for various thermodynamic quantities for such black holes. In fact, this can be further used for analysing 
  phase transition in such systems.
The phase transitions has also been studied in  large AdS black hole \cite{o}.
In fact,  there is a close analogy between the charged AdS black holes and Van der Waals fluids \cite{w, w1}. This analogy
becomes even more clear in the extended phase space \cite{w2, w4}. This is because in this case the cosmological constant can be treated
as a thermodynamical pressure. This can be used to write the first law of thermodynamics in terms of the cosmological constant.
Finally, it is also possible to define a thermodynamic volume using this first law of thermodynamics.
It will be interest to see how this analysis changes when thermal fluctuations are taken into account.
In this paper, we will use the leading order form for the corrections to the entropy  of a charged AdS  black hole to
analyze the phase transition of this system. This will be done by first calculating the corrections to the entropy of
a charged black hole explicitly, and then using this to calculate other thermodynamical quantities in the system. Finally,
that will be used for analyzing the phase transition of this system.

\section{Charged AdS Black Hole}
In this section, we will consider a charged AdS black hole.
The action for a four dimensional asymptotically AdS spacetime coupled to Maxwells
equations  can be written as
\begin{equation}
I = \int d^4 x \sqrt{-g} \left(R + \frac{6}{l}  - \frac{1}{4} F^{\mu\nu} F_{\mu\nu}\right),
\end{equation}
where $l$ is the AdS radius, and $F_{\mu\nu} = \nabla_\nu A_\mu - \nabla_\mu A_\nu $.
The field equations  for this system can now be written as
$
 G_{\mu\nu} =  3  g_{\mu\nu}/l +   F^{\mu\tau} F_\nu^{\tau}/2  -   g_{\mu\nu} F^{\tau \rho} F_{\tau \rho}/8,
$
and
$
 \sqrt{-g} \nabla_\nu F^{\mu\nu} = 0$.
The metric for this a charged AdS black hole,  can now be written as
\begin{equation}\label{c1}
ds^{2}=-f(r)dt^{2}+\frac{dr^{2}}{f(r)}+r^{2}d\Omega_{k}^{2},
\end{equation}
where
\begin{equation}\label{c2}
d\Omega_{k}^{2}=d\theta^{2}+\frac{1}{k}\sin^{2}(\sqrt{k}\theta)d\varphi^{2},
\end{equation}
and
\begin{equation}\label{c3}
f=k-\frac{2M}{r}+\frac{Q^{2}}{r^{2}}+\frac{r^{2}}{l^{2}}.
\end{equation}
The black horizon radius can be obtained by taking the real positive root of the following equation,
\begin{equation}\label{c4}
r_{+}^{4}+l^{2}r_{+}^{2}-2Ml^{2}r_{+}+l^{2}Q^{2}=0.
\end{equation}

In the path integral formalism, it is possible to calculate the amplitude for a field configuration    to 
propagate to another field configuration. This  can be done using a formalism called the Euclidean quantum gravity, 
where  the temporal coordinate is rotated on  the complex plane. Thus,  
  the partition function for the charged AdS spacetime can be written as  \cite{o, 01ab, 02ab, 04ab, 05ab}
\begin{equation}
Z = \int D g  D A \exp - I,
\end{equation}
where $I \to -i I$ is the Euclidean action for this system.
It may be noted that this can be related to the statistical mechanical partition function  \cite{hawk, hawk1}
\begin{equation}
Z = \int_0^\infty  dE \, \,  \rho (E) e^{-\beta E},
\end{equation}
where $\beta$ is the inverse of the temperature.
Now, the density of states can be written as
\begin{eqnarray}
\rho (E) = \frac{1}{2 \pi i} \int^{\beta_0+ i\infty}_{\beta_0 - i\infty}  d \beta \, \, e ^{S(\beta)} ,
\end{eqnarray}
where
\begin{equation}
S = \beta  E   + \ln Z.
\end{equation}
Usually this entropy is measured around the equilibrium temperature $\beta_0$, and all thermal fluctuations are 
neglected. 
The expression for entropy, when all the thermal fluctuations are neglected is given by 
\begin{equation}\label{T2}
S_{0}=\pi r_{+}^{2}.
\end{equation}
Similarly, the temperature of the
charged AdS black hole can be written as
\begin{equation}\label{T1}
T=\frac{l^{2}(r_{+}^{2}-Q^{2})+3r_{+}^{4}}{4\pi l^{2} r_{+}^{3}}.
\end{equation}
It is also possible define a thermodynamic volume,  in absence of thermal fluctuations as \cite{4a}
\begin{equation}\label{T3}
V=\frac{4}{3}\pi r_{+}^{3}.
\end{equation}

However, it is possible to take these thermal fluctuations into account, and 
  expand  $S(\beta)$ around the equilibrium temperature $\beta_0$ \cite{l1, SPR}, 
\begin{equation}\label{fluc}
S = S_0 + \frac{1}{2}(\beta - \beta_0)^2 \left(\frac{\partial^2 S(\beta)}{\partial \beta^2 }\right)_{\beta = \beta_0}, 
\end{equation}
where we have 
  neglected higher order corrections to the entropy. Now we can   write the density of states as
\begin{eqnarray}
\rho (E) = \frac{e^{S_0}}{ 2 \pi i}  \int^{\beta_0+ i\infty}_{\beta_0 - i\infty}  d \beta \, \,  \exp \left( \frac{1}{2}
 (\beta- \beta_0)^2 \left(\frac{\partial^2 S(\beta)}{\partial \beta^2 }\right)_{\beta = \beta_0}   \right).
\end{eqnarray}
After changing the variables, this expression can be written as
\begin{equation}
\rho(E) = \frac{e ^{S_{0}}}{\sqrt{2 \pi }} \left[\left(\frac{\partial^2 S(\beta)}{\partial \beta^2 }\right)_{\beta = \beta_0}\right]^{1/2}.
\end{equation}
Thus, we can write
\begin{equation}
S = S_0 -\frac{1}{2} \ln \left[\left(\frac{\partial^2 S(\beta)}{\partial \beta^2 }\right)_{\beta = \beta_0}\right]^{1/2}.
\end{equation}
It may be noted that this second derivative of entropy is actually a fluctuation squared of the energy. 
It is possible to  simplify this expression by using the fact that  the    microscopic degrees of freedom of a black hole can be calculated 
from a  conformal field theory \cite{card}. Thus,   the entropy can be assumed to have the  form, $S = a \beta^m  + b \beta^{-n }$, 
where $  a, b, m, n > 0$   \cite{ca}. This has an extremum at the value $\beta_0 = (n b/m a)^{1/(m+n)}$, and so expanding $S$ around this extremum, 
we can   demonstrate that \cite{l1, SPR} 
\begin{equation}
 \left(\frac{\partial^2 S(\beta)}{\partial \beta^2 }\right)_{\beta = \beta_0}  = S_0 \beta_0^{-2}.   
\end{equation}
Thus, we will use the corrected form for
the  entropy, neglecting higher order corrections can be written as 
\begin{equation} 
S=S_{0}-\frac{1}{2}\ln{S_{0}T^{2}}, 
\end{equation} 
where we have used $\beta_0 = T^{-1}$.       It may be noted that the  quantum 
fluctuations in the geometry of a black hole 
give rise to thermal fluctuations in the thermodynamics of a black hole. 
So, this correction term  will only contribute  sufficiently when  the black hole is small and 
its temperature is large. This is because we can neglect the quantum fluctuations to the geometry of a large  black 
hole. In fact, as the  thermal fluctuations only become sigfinicant for objects with large temperature, and the 
temperature of a black hole increases as its size reduces, we expect this correction term to 
only contribute 
when the size of the black hole is sufficiently small. 
In this section, we analyzed some basic aspects of thermodynamics of the charged AdS black hole.
In the next section,
we will explicitly calculate the corrections to these thermodynamic quantities due to thermal fluctuations.

\section{Phase Transition }
In this section, we will use the form for the corrections to the entropy of a charged AdS black hole to obtain
explicit expression for various thermodynamic quantities.
Then, we will use these explicit values to study phase transition in this system.   Thus, we will use the corrected form for
the  entropy, neglecting higher order corrections 
\begin{equation} 
S=S_{0}-\frac{\alpha}{2}\ln{S_{0}T^{2}}, 
\end{equation}
where we have  added the parameter $\alpha$, to help track corrections coming from the 
thermal fluctuations. So, by setting $\alpha=0$, we will recovers expression for the entropy
without any corrections. Furthermore, if we can set $\alpha=1$, 
we obtain the corrections due to thermal fluctuations.   Thus, for large black holes whose temperature is 
 very  small, we can take the limit  $\alpha\rightarrow0$, and for  small  black holes whose   
temperature sufficiently large, we can take the limit $\alpha\rightarrow1$. We will now analyze the effect of this 
parameter $\alpha$ on the thermodynamical stability of a black hole. Now using temperature and entropy given by Eqs.  
(\ref{T1}) and (\ref{T2}),  we can obtain the following expression,
\begin{equation}\label{L2}
S=\pi r_{+}^{2}-\alpha\ln[l^{2}(r_{+}^{2}-Q^{2})+3r_{+}^{4}]+\alpha\ln[4\sqrt{\pi}l^{2}r_{+}^{2}].
\end{equation}
It is clear that the effect of logarithmic correction is reduction of the entropy.
Specific heat at constant pressure and constant volume can be obtained  as follows,
\begin{equation}\label{L3}
C_{p}=\frac{4}{3}\frac{9\pi r_{+}^{6}+(2\pi l^{2}-6\alpha)r_{+}^{4}-\pi l^{2}Q^{2}r_{+}^{2}}{3Q^{2}l^{2}-l^{2}r_{+}^{2}+3r_{+}^{4}},
\end{equation}
and
\begin{equation}\label{L4}
C_{v}=\frac{6\pi r_{+}^{6}+(2\pi l^{2}-6\alpha)r_{+}^{4}-2\pi l^{2}Q^{2}r_{+}^{2}-2\alpha Q^{2}l^{2}}{3Q^{2}l^{2}-l^{2}r_{+}^{2}+3r_{+}^{4}}.
\end{equation}
We will use $C_{v}\geq0$  to investigate phase transition. Ratio of above quantities is denotes by,
\begin{eqnarray}\label{L5}
\gamma &=&\frac{2}{3}\frac{9\pi r_{+}^{6}+(2\pi l^{2}-6\alpha)r_{+}^{4}-\pi l^{2}Q^{2}r_{+}^{2}}{3\pi r_{+}^{6}+(\pi l^{2}-3\alpha)r_{+}^{4}-\pi l^{2}Q^{2}r_{+}^{2}-\alpha Q^{2}l^{2}}\nonumber\\
&=&\frac{2}{3}\left[1+\frac{6\pi r_{+}^{6}+(\pi l^{2}-3\alpha)r_{+}^{4}+\alpha l^{2}Q^{2}}{3\pi r_{+}^{6}+(\pi l^{2}-3\alpha)r_{+}^{4}-\pi l^{2}Q^{2}r_{+}^{2}-\alpha Q^{2}l^{2}}\right].
\end{eqnarray}
It is clear that if we have,
\begin{equation}\label{L6}
6\pi r_{+}^{6}+(\pi l^{2}-3\alpha)r_{+}^{4}+\alpha l^{2}Q^{2}=0,
\end{equation}
then $\gamma=\frac{2}{3}$, which is smaller than the cases of ideal classic gas ($\gamma=\frac{5}{3}$)
and extreme relativistic gas ($\gamma=\frac{4}{3}$).
In the case of $\alpha\rightarrow0$, the above condition reduced to $r_{+}^{2}=-\frac{l^{2}}{6}$,  which seems illegible.
Therefore, existence of logarithmic correction is necessary to have $\gamma=\frac{2}{3}$.
We can obtain the following expression for total pressure,
\begin{equation}\label{L7}
P=\frac{3\pi r_{+}^{6}+(\pi l^{2}-3\alpha)r_{+}^{4}-\pi l^{2}Q^{2}r_{+}^{2}-\alpha Q^{2}l^{2}}{8\pi^{2}l^{2} r_{+}^{6}}.
\end{equation}
It may be noted that this is a   decreasing function of $\alpha$.
Now we using   Eqs.  (\ref{T1}), (\ref{T3}) and (\ref{L7}),
we obtain  $PV\propto T$, and it is  satisfied at $\alpha\rightarrow0$ or $Q\rightarrow0$. So, for the infinitesimal $\alpha Q$,  we can write,
\begin{equation}\label{L8}
PV=\frac{2}{3}\pi T.
\end{equation}
Then, we can find enthalpy as follow,
\begin{equation}\label{L9}
H=\frac{3\pi r_{+}^{4}+(2\pi l^{2}-6\alpha)r_{+}^{2}+Q^{2}l^{2}}{3\pi l^{2} r_{+}},
\end{equation}
which can be considered as black hole mass $M$ \cite{Dolan1,JJP}. Hence we can use the first law of thermodynamics \cite{Dolan2},
\begin{equation}\label{L10}
dM=TdS+VdP+\Phi dQ,
\end{equation}
to extract electric potential $\Phi=dM/dQ=dH/dQ$. It is easy to find,
\begin{equation}\label{L11}
\Phi=\frac{3\alpha l^{2}Q^{2}-2\alpha Ml^{2}r_{+}+l^{2}(3\pi Q^{2}+2\alpha)r_{+}^{2}-(3\pi l^{2}-13\alpha)r_{+}^{4}-9\pi
r_{+}^{6}}{6\pi r_{+}^{4}(2r_{+}^{3}+l^{2}r_{+}-Ml^{2})}Q,
\end{equation}
which is increasing function of $\alpha$ out of inner horizon, while it is decreasing function of $\alpha$ in the center of black hole. It means that the effect of corrected terms is increasing the electric potential at the horizon. On the other hand, in the center of the black hole, logarithmic corrections decreased value of the electric potential.\\

Finally,  we can find  Helmholtz and Gibbs energies as follows,
\begin{eqnarray}\label{L13}
F=&-&\frac{3\pi r_{+}^{6}-(3\pi l^{2}-18\alpha)r_{+}^{4}-9\pi l^{2}Q^{2}r_{+}^{2}-2\alpha Q^{2}l^{2}}{12\pi l^{2} r_{+}^{3}}\nonumber\\
&-&\frac{\alpha(l^{2}Q^{2}-l^{2}r_{+}^{2}-3r_{+}^{4})}{4\pi l^{2}r_{+}^{3}}\ln(3r_{+}^{4}+l^{2}r_{+}^{2}-l^{2}Q^{2})\nonumber\\
&+&\frac{\alpha(l^{2}Q^{2}-l^{2}r_{+}^{2}-3r_{+}^{4})}{4\pi l^{2}r_{+}^{3}}\ln(4\sqrt{\pi}l^{2}r_{+}^{2}),
\end{eqnarray}
and
\begin{eqnarray}\label{L14}
G=&+&\frac{3\pi r_{+}^{6}+(5\pi l^{2}-24\alpha)r_{+}^{4}+7\pi l^{2}Q^{2}r_{+}^{2}}{12\pi l^{2} r_{+}^{3}}\nonumber\\
&-&\frac{\alpha(l^{2}Q^{2}-l^{2}r_{+}^{2}-3r_{+}^{4})}{4\pi l^{2}r_{+}^{3}}\ln(3r_{+}^{4}+l^{2}r_{+}^{2}-l^{2}Q^{2})\nonumber\\
&+&\frac{\alpha(l^{2}Q^{2}-l^{2}r_{+}^{2}-3r_{+}^{4})}{4\pi l^{2}r_{+}^{3}}\ln(4\sqrt{\pi}l^{2}r_{+}^{2}).
\end{eqnarray}
All of these quantities are decreasing function of $\alpha$, which means the logarithmic correction term decreases the energy.
\begin{equation}\label{L12}
E=\frac{3\pi r_{+}^{6}+(3\pi l^{2}-9\alpha)r_{+}^{4}+3\pi l^{2}Q^{2}r_{+}^{2}+\alpha Q^{2}l^{2}}{6\pi l^{2} r_{+}^{3}},
\end{equation}
Thus, the energy of the system is decreases by thermal fluctuations.\\

\begin{figure}[th]
\begin{center}
\includegraphics[scale=.5]{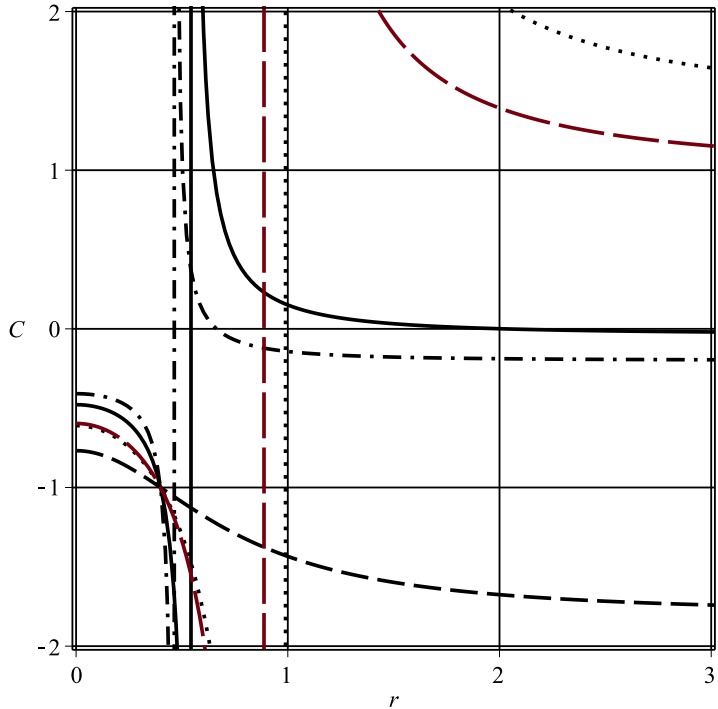}
\caption{Specific heat in constant volume versus horizon radius. $\alpha=0$ (dashed line), $\alpha=0.5$ (solid line), $\alpha=0.8$ (dash dotted line), $\alpha=1$ (Red long dashed line), $\alpha=1.6$ (dotted line)}
\label{fig1}
\end{center}
\end{figure}

Now using the specific heat given by the equation (\ref{L4}),  we can investigate phase transition of the black hole. The specific
heat can be written in terms of the horizon radius as a parameter, by using Eqs.  (\ref{c4}), (\ref{L9}) and (\ref{L4})
\begin{equation}\label{P1}
C_{v}=\frac{6\pi(2\alpha-1) r_{+}^{2}-\pi+12\alpha}{6\pi r_{+}^{2}+\pi-18\alpha},
\end{equation}
where $l=1$ assumed. An important consequence of logarithmic correction
is that the black hole is unstable for $\alpha=0$ and   stable for $\alpha\geq\frac{1}{2}$.
In the Fig. \ref{fig1},  we show behavior of the specific heat given by the equation (\ref{P1}), for several values of $\alpha$.
We can see that there are some stable regions $r_{+}\geq r_{m}$ corresponding to $\alpha\geq\frac{1}{2}$. Furthermore, $r_{+}=r_{m}$ is minimum
value of the black hole horizon radius where phase transition happen. For example, at $\alpha=1.6$ phase transition happen at $r_{+}=1$.
So,  the black hole is completely stable for $r_{+}\geq1$, for instance $Q=1$ and $M=2$ gives stable black hole with $r_{+}\approx1.25$.

\section{Conclusions}
In this paper, we analyzed  the effects of thermal fluctuations on a charged AdS black hole.
This was done by analyzing the corrections to thermodynamics of this  charged AdS black hole due to thermal fluctuations.
It was also demonstrated the the logarithmic term is the leading order correction term to  the entropy of this charged AdS black
hole. We also calculated other corrections to other important thermodynamic quantities for this black hole.
Finally, we used the corrected value of the specific heat to analyze the phase transition in this black hole.
The corrections terms to the entropy may be related to the conformal field theory description of all black solutions \cite{conf}.
It may be noted that asymptotically  AdS space-time have been studied because of their
importance in the AdS/CFT correspondence ~\cite{Maldacena:1997re}.
According to the AdS/CFT correspondence, the string theory on AdS is dual to a conformal field theory on the boundary of that AdS.
It has also been demonstrated using the AdS/CFT correspondence that the asymptotically AdS black hole is dual to
a strongly coupled gauge theory at finite temperature \cite{as, as11, as22, as44}.
It is possible to analyze the strongly correlated condensed matter physics using
AdS/CFT correspondence. In fact, a holographic model of
superconductors has also been constructed from black hole
solutions using the AdS/CFT correspondence ~\cite{Hartnoll:2008kx}. It will be interesting
to analyze the CFT dual to thermal fluctuations on the AdS, and discuss its possible applications
for the strongly correlated condensed matter systems.

\end{document}